\RequirePackage{fix-cm}
\RequirePackage{fixltx2e}
\documentclass[english]{article}
\usepackage[T1]{fontenc}
\usepackage[latin9]{inputenc}
\usepackage{geometry}
\usepackage[active]{srcltx}
\usepackage{color}
\usepackage{babel}
\usepackage{array}
\usepackage{float}
\usepackage{multirow}
\usepackage{amsmath}
\usepackage{amsthm}
\usepackage{amssymb}
\usepackage{stmaryrd}
\usepackage{graphicx}
\usepackage[authoryear]{natbib}
\usepackage[unicode=true,pdfusetitle,
 bookmarks=true,bookmarksnumbered=true,bookmarksopen=true,bookmarksopenlevel=1,
 breaklinks=true,pdfborder={0 0 1},backref=false,colorlinks=true]
 {hyperref}
\hypersetup{
 linkcolor=blue, citecolor=blue, urlcolor=blue, filecolor=blue,pdfpagelayout=OneColumn, pdfnewwindow=true,pdfstartview=XYZ, plainpages=false, pdfpagelabels, hyperindex=true}

\makeatletter

\providecommand{\tabularnewline}{\\}

\numberwithin{equation}{section}
\numberwithin{table}{section}

\usepackage{lastpage}
\usepackage{amsmath}
\usepackage{graphicx}
\usepackage{color}
\usepackage{fancybox} 
\usepackage{moreverb} 
\usepackage{listings} 
\usepackage{diagbox}
\usepackage{array}
\usepackage{makecell}
\usepackage{colortbl}
\usepackage{prettyref}
\usepackage{textcomp}
\usepackage[title]{appendix}
\usepackage[font=footnotesize,labelfont=bf]{caption}
\usepackage{authblk}
\usepackage{indentfirst}
\usepackage{ifpdf} 
\ifpdf 
 \IfFileExists{lmodern.sty}{\usepackage{lmodern}}{}

\fi 

\let\myTOC\tableofcontents
\renewcommand\tableofcontents{%
  \pdfbookmark[1]{\contentsname}{}
  \myTOC
}

\def\LyX{\texorpdfstring{%
  L\kern-.1667em\lower.25em\hbox{Y}\kern-.125emX\@}
  {LyX}}

\@addtoreset{footnote}{section}

\usepackage{dsfont}
\usepackage{bbm}
\usepackage{lineno}


\author[1,2]{Wen Chen}
\author[1]{Nicolas Langren\'e}
\affil[1]{\small Commonwealth Scientific and Industrial Research Organisation, Data61, RiskLab Australia}
\affil[2]{\small Corresponding author. Email: \href{wen.chen@csiro.au}{wen.chen@csiro.au}}

\makeatother

\begin{document}
\title{Deep neural network for optimal retirement consumption\\
in defined contribution pension system}
\maketitle
\begin{abstract}
In this paper, we develop a deep neural network approach to solve
a lifetime expected mortality-weighted utility-based model for optimal
consumption in the decumulation phase of a defined contribution pension
system. We formulate this problem as a multi-period finite-horizon
stochastic control problem and train a deep neural network policy
representing consumption decisions. The optimal consumption policy
is determined by personal information about the retiree such as age,
wealth, risk aversion and bequest motive, as well as a series of economic
and financial variables including inflation rates and asset returns
jointly simulated from a proposed seven-factor economic scenario generator
calibrated from market data. We use the Australian pension system
as an example, with consideration of the government-funded means-tested
Age Pension and other practical aspects such as fund management fees.
The key findings from our numerical tests are as follows. First, our
deep neural network optimal consumption policy, which adapts to changes
in market conditions, outperforms deterministic drawdown rules proposed
in the literature. Moreover, the out-of-sample outperformance ratios
increase as the number of training iterations increases, eventually
reaching outperformance on all testing scenarios after less than 10
minutes of training. Second, a sensitivity analysis is performed to
reveal how risk aversion and bequest motives change the consumption
over a retiree's lifetime under this utility framework. Our results
show that stronger risk aversion generates a flatter consumption pattern;
however, there is not much difference in consumption with or without
bequest until age 103. Third, we provide the optimal consumption rate
with different starting wealth balances. We observe that optimal consumption
rates are not proportional to initial wealth due to the Age Pension\textit{
}payment. Forth, with the same initial wealth balance and utility
parameter settings, the optimal consumption level is different between
males and females due to gender differences in mortality. Specifically,
the optimal consumption level is slightly lower for females~until~age~84.

\vspace{-1mm}
$\,$

\vspace{-1mm}

$\,$

\vspace{-1mm}

$\,$
\end{abstract}

\section{Introduction\label{sec:introduction}}

The global trend of transition from defined benefit (DB) to defined
contribution (DC) system means that responsibility switched from employers
to employees, making life-cycle management an important and relevant
topic to every individuals. In Australia, the DC system established
in 1992, requires employers to contribute a minimum percentage (Superannuation
Guarantee rate\footnote{The SG rate was 9.5\% on 1 July 2014, and will remain 9.5\% rate until
30 June 2021, and is set to have five annual increases, where the
SG rate will increase to 12\% by July 2025.}) of employee's earnings to a superannuation fund. Over 28 years,
the Australia DC system has reached a mature stage, and achieved retirement
savings of around $\$3$ trillion (AUD) in asset, and has the highest
proportion in DC assets ($86\%$) relative to DB assets ($14\%$)
\citep{Willis2020}. More new retirees now own significant superannuation
savings accumulated during their working lives. However, there is
little guidance on how to achieve the best strategies after retirement.
Also due to reasons such as fear of ruin and bequest motive, most
retirees tend to only withdraw the statutory minimum rate from their
retirement account (\citealt{Sneddon2016}, \citealt{Deravin2019}),
thus not optimising benefit from their retirement savings. Designing
income solution becomes a major challenge for the Australian superannuation
fund industry. The retirement income solution in the decumulation
phase is underdeveloped. In the academic as well as the industrial
practitioner literature, the utility framework has been used to analyse
complex life-cycle management problems which involve consumption,
asset allocation and possibly other decisions. The utility function
reduces the dimension of the problem by providing a quantitative measure
of uncertain retirement outcome satisfaction resulting from different
deterministic or dynamic financial strategies, such as drawdown strategies
and investment strategies. Since the seminal contribution of \citet{Yaari1964,Yaari1965}
on utility based life-cycle models in asset allocation and optimal
consumption, alternative methods and extensions have been investigated.

For example, \citet{thorp2007} investigated optimal investment and
annuitisation strategies for the decumulation phase using a Hyperbolic
Absolute Risk Aversion (HARA) utility function. \citet{Huang2012}
proposed a partial differential equation (PDE) framework for optimal
consumption with stochastic force of mortality and deterministic investment
returns to maximise discounted expected Constant Relative Risk Aversion
(CRRA) utility over a random time horizon. \citet{Mao2014} also looked
at the optimal consumption problem during decumulation, in conjunction
with related decision problems such as optimal retirement age and
optimal leisure time, using CRRA utility. \citet{ding2014dynamic}
used a CRRA utility, focusing on optimal portfolio allocation and
the effect of bequest. \citet{Andreasson2017} investigated the optimal
consumption, investment and housing decisions with means-tested public
pension in retirement, using a HARA utility function discounted by
inflation. \citet{Butt2018} considered the optimal consumption and
asset allocation problems with Age Pension taken into account, using
CRRA utility. \citet{Deravin2019} designed a mortality weighted CRRA
utility-based metric called the Members Default Utility Function (MDUF)
to assist industry to design retirement outcome solutions for the
decumulation phase. Finally, a recent attempt to avoid the expected
utility approach is \citet{Forsyth2019}, in which a CVaR approach
is used to manage depletion risk via asset allocation throughout both
the accumulation and decumulation phases in a DC system.

In practice, the most popular approach for solving such optimal decumulation
stochastic control problems is numerical dynamic programming with
state discretisation and interpolation (\citealt{rust1996numerical},
\citealt{Andreasson2017}, \citealt{Butt2018}, \citealt{Deravin2019},
\citealt{Jin2020}). Other popular approaches include analytical or
semi-analytical solutions (\citealt{Yaari1965}, \citealt{ding2014dynamic})
and numerical schemes for Hamilton-Jacobi-Bellman (HJB) PDE (\citealt{Cairns2006},
\citealt{Huang2012}, \citealt{Forsyth2019}).

The analytical solution approach is the most attractive, but is unfortunately
usually infeasible or requires too significant simplifications of
the original problem to remain useful in practice. The numerical dynamic
programming and PDE approaches are able to solve more realistic problem
with a greater set of features, but both suffer from the curse of
dimensionality, which puts a limit on the number of stochastic factors
which can be accounted for.

In order to break the limitations of these classical numerical approaches,
we propose in this paper a deep neural network approach (DNNs, \citealt{Goodfellow2016}).
The principle is to model the unknown consumption policy by a DNN
and to optimise it directly by modern gradient descent techniques.
This bypasses the dynamic programming principle altogether, overcomes
the curse of dimensionality (\citealt{poggio2017}, \citealt{han2018solving}),
and makes it possible to consider as many realistic features and stochastic
factors as deemed necessary to obtain informative consumption advice
in the decumulation phase in a DC system. Being able to account for
a great range of customised features and personal objectives in the
decision-making process with machine learning technology shall improve
member engagement in the DC system \citep{Fry2019}.

Approximating policies by DNNs have been well explored in the reinforcement
learning field, see \citet[Chapter 13]{Sutton2018}, with great practical
success as illustrated for example by the famous AlphaGo and AlphaGo
Zero programs \citet[Chapter 16]{Sutton2018}. For the most part,
reinforcement learning is concerned with infinite time horizon problems,
coupled with discrete-valued control spaces. By contrast, this paper
addresses a finite time horizon stochastic control problem (due to
mortality) with continuous policies (namely consumption). In this
context, optimal policies should explicitly depend on time, either
time since inception $t$ or time to maturity $T-t$, where $T$ is
the maturity.

Classically, the optimal policies of stochastic control problems with
finite time horizon are estimated in a backward manner, taking advantage
of the dynamic programming principle \citep{Beckmann1968}. Recently,
DNNs have been used in this context in \citet{Hure2018}, \citet{Bachouch2018}
and \citet{Fecamp2019} for approximating value functions, policy
functions or both. As an alternative to the classical backward dynamic
programming approach, \citet{Han2016} suggested to use one DNN policy
function for each decision time $t_{i}$, yielding a collection of
sub-networks, indexed by time, and trained simultaneously. This global
policy learning approach has been used in \citet{Fecamp2019} for
financial option hedging problems, in \citet{Guo2019} for robust
portfolio allocation, and has been adapted to optimal stopping problems
in \citet{Becker2019} with application to exotic option pricing.

One modification of this global approach is to consider one single
DNN and include time as part of the inputs, in addition to all the
other state variables. The very same DNN can then be used at all decision
times. \citet{Fecamp2019} also implemented and tested this modification,
and \citet{Li2019} used it for an asset allocation problem. This
single DNN approach can be thought of as the adaptation of recurrent
neural networks (RNN) to finite-horizon problems, for which policies
do change over time and therefore time needs to be part of the inputs
of this single DNN. As such a modification is much more parsimonious,
easier to train and better suited to problems for which policies vary
in a smooth manner over time, we make use of this approach in this
paper to model and learn optimal consumption policies.

The main contribution of this paper is the proposed use of a DNN-policy
approach to realistic utility maximisation problems driven by multi-factor
economic stochastic model. Our approach is very general: as it does
not rely on dynamic programming, the objective function is not limited
to expected utility and is allowed to be more sophisticated an personalised;
as it is simulation-based, all the realistic features of the problem,
such as fees and Age Pension policy can be easily accounted for; finally,
the ability of DNNs to handle large-scale problems means that the
economic scenario generator (ESG) used to represent the state of the
economy can contain as many stochastic variables as deemed necessary.
The second contribution of this paper is a proposed 7-factor ESG which
we use to perform numerical experiments and obtain findings on realistic
decumulation test cases in the Australian DC pension system.

The paper is organised as follows. Section \ref{sec:problem_formulation}
formulates the problem and introduces useful notations. Section \ref{sec:ESG}
introduces the ESG used for our numerical tests. Section \ref{sec:DNN}
details the DNN numerical approach used to solve our utility maximisation
problem. Section \ref{sec:numerical_results} presents our numerical
results on several test cases based on the Australian superannuation
system. Finally, Section \ref{sec:conclusion} concludes the paper
and suggests several areas of future work.

\section{Problem formulation\label{sec:problem_formulation}}

In this section, we introduce our utility-based objective function
(subsection \ref{subsec:objective-function}), a stochastic economic
model in the form of an Economic Scenario Generator (ESG, subsection
\ref{sec:ESG}) which provides us with Monte Carlo simulations of
economic variables, means-tested Age Pension (subsection \ref{subsec:age-pension-simulation})
and wealth dynamics (subsection \ref{subsec:wealth}).

\subsection{Objective function\label{subsec:objective-function}}

The objective of this problem is to maximise the mortality-weighted
expected utility through a dynamic consumption policy in order to
obtain the optimal consumption level in the future under any market
conditions. We formulate the objective function similar to the MDUF
proposed in \citet{Bell2017,Deravin2019}, with the wealth of the
retiree following the dynamics defined in subsection \ref{subsec:wealth}.

When entering retirement at time $t=0$, the total utility of one
single person is defined as follows.
\begin{equation}
V_{0}\left(w_{0}\right)=\max_{\{c_{t}\}_{0\le t\le T}}\mathbb{E}\left[\sum_{t=0}^{T}\left\{ _{t}p_{x}u\left(c_{t}\right)+_{t-1|}q_{x}v\left(w_{t}\right)\right\} \right]\label{eq:objective}
\end{equation}
subject to
\begin{equation}
c_{t}\in\left[0,w_{t}+a_{t}\right]\label{eq:consumption_constraint}
\end{equation}
where $x$ is the retirement age (in years) which is set to be $67$,
and the maximum age is set to be $108$, therefore the time horizon
is $T=41$. $c_{t}$ is the annual consumption, $w_{t}$ is the wealth
and $a_{t}$ is the Age Pension payment subjected to means test at
decision time $t$, where $t=0,1,...,T$. All these three variables
are expressed in real terms, adjusted for inflation, when computing
the utilities. The consumption $c_{t}$ should be positive and less
than the total wealth plus the Age Pension payment at any time $t$.
The consumption utility function $u\left(c_{t}\right)$ is a CRRA
type with consumption risk aversion parameter $\rho$ ($\rho=0$ being
risk neutral) defined as
\begin{equation}
u\left(c_{t}\right)=\frac{c_{t}^{1-\rho}}{1-\rho}\,,\label{uct}
\end{equation}
and the final utility of the residual wealth $v\left(w_{t}\right)$
if the person dies between $t-1$ and $t$ is defined as
\begin{align}
v\left(w_{t}\right) & =\frac{w_{t}^{1-\rho}}{1-\rho}\left(\frac{\phi}{1-\phi}\right)^{\rho}\,,\label{vbt}
\end{align}
with strength of bequest motive $\phi\in[0,1)$. The higher $\phi$,
the stronger the bequest motive. Conversely, $\phi=0$ means there
is no desire to leave wealth unspent after death.

$_{t}p_{x}$ is the probability of surviving at age $x+t$ conditional
on being alive at age $x$. It can approximate the state of health.
$_{t-1|}q_{x}$ is the probability of death during $(x+t-1,x+t]$
conditional on being alive at age $x$. We assume that the male and
female retirees are subject to the mortality rates in Australian Life
Tables 2015-17 published by Australian Government Actuary (AGA, \citealt{AGA2017}).
Since the problem involves ageing and the time span covers several
decades, it is important to allow for future improvements in mortality
rates. We assume the mortality rates decrease with the 25-year improvement
factor as published by the AGA, to project the mortality rates that
could be expected to occur over an individual's lifetime\footnote{The details on how to use life tables and the mathematical form of
incorporating future improvements can be found at \url{http://aga.gov.au/publications/life\_table\_2010-12/07-Part3.asp}}. We also assume that mortality rates are independent of retirement
wealth, consumption and portfolio returns.

\subsection{Economic scenario generator\label{sec:ESG}}

To complete the definition of the utility maximisation problem \eqref{eq:objective},
we need a model to describe economic variables such as inflation and
the asset returns which affect the wealth of the retiree as well as
its age pension entitlements. In \citet{Bell2017,Deravin2019}, the
authors used a fixed risk-free rate and assume the returns of the
risky asset follow a normal distribution. Such constant or fixed investment
return distribution assumptions are common and convenient for computational
reasons, but can be deemed too basic to properly capture uncertainties
inherent in the financial market for retirement management. In the
present work, one convenient consequence of the proposed deep learning
approach for solving \eqref{eq:objective} (Section \ref{sec:DNN})
is that the specific choice of model for the economic variables of
the problem does not restrict the numerical feasibility of the utility
maximisation problem \eqref{eq:objective}.

In this paper, we propose to use a multivariate stochastic investment
model in the form of an Economic Scenario Generator (ESG, \citealt{moudiki2016economic}),
which are commonly applied in the actuarial field to simulate future
economic and financial variables. A well-known ESG is Wilkie's four-factor
investment model \citep{Wilkie1984} which models inflation rates,
equity returns and bond returns as stochastic time series through
a cascading structure to describe the investment returns. A series
of updates and extensions of this model have been proposed on both
practical and theoretical aspects \citep{Wilkie1995,Sahin2008,Wilkie2017a,Wilkie2019}.
The Ahlgrim model \citep{Ahlgrim2005} developed for the Casualty
Actuarial Society extends the model to include real estate prices.
\citet{Zhang2018updating} revisit Wilkie's model and examine the
model performance for the United States. Based on the model in \citet{Wilkie1995},
\citet{Butt2012} propose an ESG model to investigate investment strategies
for the Australian DC pension system. \citet{Chen2020ESG} extended
the SUPA model of \citet{Sneddon2016} to a 14-factor SUPA model for
the Australian system, and use it to quantify uncertainty and model
downside risks with retirement savings in the accumulation and decumulation
phases.

\subsubsection{Seven-factor economic scenario generator}

We design a seven-factor ESG covering inflation and six asset classes,
which is sufficient for completing the description of the optimal
decumulation problem \eqref{eq:objective}.

Our ESG models the dynamics of inflation $q(t)$, risky asset returns
such as domestic and international total equity returns $e(t)$ and
$n(t)$, real estate returns $h(t)$, and defensive asset returns
such as interest rates $s(t)$ and domestic and international bond
returns $b(t)$ and $o(t)$. Similar to Wilkie's model \citep{Wilkie1984},
we assume the inflation rate $q(t)$ follows a discretised mean-reverting
Ornstein-Uhlenbeck process (a.k.a. AR(1) process). From there, the
specific dynamics of each economic factor and their dependence are
summarised in Table \ref{table:calibration} in Appendix \ref{sec:Calibration}.
The model is calibrated to historical data from year 1992 when the
Superannuation Guarantee started, to year 2020. The data are obtained
from the Reserve Bank of Australia (RBA), Australian Bureau of Statistics
(ABS) and Bloomberg. Table \ref{table:calibration} also reports the
calibrated values on the considered dataset.

The simulated inflation allows us to project the future Age Pension
rates and the mean-test thresholds values, so we can compute the future
Age Pension eligible payment under the current pension policy. We
also use inflation-adjusted consumption and wealth to compute their
utilities. The simulated asset returns allow us to compute the returns
of the predefined portfolios described in the next subsection \ref{subsec:Portfolio}.

\subsubsection{Portfolio returns and fund management fees\label{subsec:Portfolio}}

We build investment portfolios across the aforementioned six different
asset classes. According to the Government website MoneySmart, there
are four common types of life-cycle investment strategies: \textit{Cash},
\textit{Conservative}, \textit{Balanced} and \textit{Growth} with
$0\%$, $30\%$, $70\%$ and $85\%$ invested in growth assets respectively.
In this paper, we adopt the portfolios settings in \citet{Chen2020}
where the growth (risky) portfolio includes $50\%$ Australian equity
(domestic) $e(t)$, $30\%$ international equity (excluding Australia)
$n(t)$ and $20\%$ property $h(t)$, and the defensive portfolio
includes $30\%$ risk-free term deposit $s(t)$, $50\%$ domestic
bond $b(t)$ and $20\%$ international bond $o(t)$\footnote{These weights reflect a lower exposure to both growth assets (50\%
v 66\%) and international assets (25\% v 36\%) on the average Australian
Prudential Regulation Authority (APRA) regulated superannuation fund
asset allocations as at December 2019, reflecting the conservative
nature of typical retiree allocations. More information is available
at \url{https://www.apra.gov.au/quarterly-superannuation-statistics} (Retrieved
June 4, 2020).}. Then the returns of the growth and defensive portfolios are given
as follows:
\begin{eqnarray*}
R_{\mathrm{growth}}(t) & = & 50\%e(t)+30\%n(t)+20\%h(t)\,,\\
R_{\mathrm{defensive}}(t) & = & 30\%s(t)+50\%b(t)+20\%o(t)\,.
\end{eqnarray*}
In our numerical examples, we select the \textit{Balanced} investment
strategy, with $\omega=70\%$ invested in growth assets and $30\%$
invested in defensive assets.
\begin{equation}
R(t)=\omega\cdot R_{\mathrm{growth}}(t)+\left(1-\omega\right)R_{\mathrm{defensive}}(t)\,.\label{eq:Rt}
\end{equation}
We keep these weights fixed throughout retirement.

In Australia, all superannuation funds charge management fees, and
the impact of fees should not be ignored when modelling retirement
consumption. \citet{PC2019} reported that ``\textit{balances are
eroded by fees and insurance}'' and that fees were ``\textit{the
biggest drain on net returns}''. The total management fee consists
of at least administration fee, super fund member fee with an indirect
cost ratio depending on the chosen fund level, and investment fee
depending on the chosen investment strategy and balance. In some superannuation
funds, the investment fee is also associated with its performance,
and the total cost could include other fees, such as advice fees,
exit fees and brokerage fees.

Total fees vary significantly by superannuation funds level, balance
and investment option. In our example, we use the rates provided by
the MoneySmart superannuation calculator of the Australian Securities
and Investments Commission (ASIC)\footnote{More details about the superannuation fund fees and cost can be found
at \url{https://www.moneysmart.gov.au/tools-and-resources/calculators-and-apps/account-based-pension-calculator}}. We choose a medium-level fund with annual administration fee of
$\$50$, indirect cost ratio of $0.6\%$ and\textit{ Balanced} investment
option with investment fee of $0.5\%$.

\subsection{Age Pension simulation\label{subsec:age-pension-simulation}}

The Age Pension is a government payment scheme which provides income
to help Australian retirees to cover their cost of living. It is paid
to people who meet the retirement age requirement, subject to an income
test and an asset test. The payment rates also depend on the family
(single or couple) and homeownership status. The pension age in 2020
is 66 and will be gradually increased to 67 by 2023. The payment rates
and the thresholds are adjusted by changes in Consumer Price Index
(CPI) or the Pensioner and Beneficiary Living Cost Index twice a year\footnote{More details about Age Pension can be found at \url{https://www.dss.gov.au/seniors/benefits-payments/age-pension}}.

In the following, we consider a 67-year-old single homeowner who converted
all liquid assets into the account-based pension and has no other
income stream. We assume the payment rates and test thresholds are
only adjusted for inflation rates annually. Based on these assumptions,
the payment is determined by the wealth, deemed income from the liquid
assets, and the compound inflation
\begin{align}
Q_{t} & =e^{\Sigma_{s=0}^{t}q_{s}}\,.\label{eq:Qt}
\end{align}
The Age Pension $A_{t}$ this person can receive at time $t$ is the
minimum of the payment under the asset test $A_{t}^{A}$ and income
test $A_{t}^{I}$, that is $A_{t}\left(W_{t},Q_{t}\right)=\min\left(A_{t}^{A},A_{t}^{I}\right)$
where
\[
A_{t}^{A}=\max\left(A_{t}^{\max}-\tau^{A}\max\left[\left(W_{t}-W_{t}^{A}\right),0\right]\right)
\]
\[
A_{t}^{I}=\max\left(A_{t}^{\max}-\tau^{I}\max\left[r_{1}\min\left(W_{t},W_{t}^{I}\right)+r_{2}\max\left(W_{t}-W_{t}^{I},0\right)-I_{t},0\right],0\right).
\]
$A_{t}^{\max}$ is the maximum Age Pension, $W_{t}^{A}$ is the asset
test threshold for full pension, $I_{t}$ is the income test cutoff
point . $W_{t}^{I}$ is the lower deeming asset threshold at time
$t$. These four variables are adjusted for inflation with $x_{t}=x_{0}Q_{t}$
for $x\in\left[A^{\max},W^{A},W^{I},I\right]$ and $t\in[0,T].$ $\tau^{A}$
and $\tau^{I}$ are the taper rates for asset and income test; $r_{1}$
and $r_{2}$ are the lower and higher deeming rates; $r_{1}<r_{2}$.
These four rates are determined by the Age Pension policy, therefore
they are assumed to be constants in this paper. The parameters are
listed in Table \ref{tab:pension}.
\begin{table}[h]
\centering{}%
\begin{tabular}{ll|ll}
\hline 
Parameter & Value & Parameter & Value\tabularnewline
\hline 
Full Age Pension $A_{t}^{\max}$ & $\$24,619$ & Income test limit $I_{t}$ & $\$4536$\tabularnewline
Asset test limit $W_{t}^{A}$ & $\$263,250$ & Lower deeming rate limit $W_{t}^{I}$ & $\$51,800$\tabularnewline
Asset taper rate $\tau^{A}$ & $0.3\%$ & Lower deeming rate $r_{1}$ & $0.25\%$\tabularnewline
Income taper rate $\tau^{I}$ & $50\%$ & Higher deeming rate $r_{2}$ & $2.25\%$\tabularnewline
\hline 
\end{tabular}\caption{Means-tested Age Pension rates published as of June 2020.\label{tab:pension}}
\end{table}

\subsection{Dynamics of wealth\label{subsec:wealth}}

In our model, the wealth is an endogenous stochastic variables which
is determined by the market changes and the consumption decisions.
The wealth at time $t+1$ is given as follows:
\begin{equation}
W_{t+1}=\left(W_{t}+A_{t}\left(W_{t},Q_{t}\right)-C{}_{t}-\mathrm{Fee}_{t}\right)e^{R_{t}}\label{eq:Wt}
\end{equation}
where $W_{t}$ , $A_{t}$ , $C_{t}$ and $\mathrm{Fee}_{t}$ are the
future (non-deflated) value of wealth, Age Pension, consumption and
fund management fees. The discounted variables $w_{t}=W_{t}/Q_{t}$,
$a_{t}\left(W_{t},Q_{t}\right)=A_{t}\left(W_{t},Q_{t}\right)/Q_{t}$
and $c_{t}=C_{t}/Q_{t}$ are the wealth, Age Pension and consumption
in real terms. Note that the total consumption consists of the drawdown
from the wealth plus the Age Pension payment. When evaluating the
utility of the future consumption and residual wealth, we used consumption
and wealth in real terms.

\section{Numerical method\label{sec:DNN}}

After the problem description, in this section we focus on the numerical
method and algorithm. We first introduce the objective function and
then the DNN control policy which is a parametric function determining
the optimal consumption at any time in any possible scenario.

\subsection{Empirical objective function}

Problem \eqref{eq:objective} is a discrete-time finite horizon stochastic
control problem. Its optimal policy (consumption) is affected in feedback
form by the stochastic state variables of the ESG described in Subsection
\ref{sec:ESG} and by the wealth (Subsection \eqref{subsec:wealth}).
More specifically, the state variables of Problem \eqref{eq:objective}
are the wealth $W$ (equation \eqref{eq:Wt}), the investment return
$R$ (equation \eqref{eq:Rt}) and the compound inflation discount
factor $Q$ (equation \eqref{eq:Qt}). We use a parametric consumption
policy approach, meaning that we model the set of possible policies
by a class of functions $(t,x)\in\mathbb{R}^{d+1}\shortrightarrow c\left(t,x;\beta\right)$
indexed by a parameter $\beta\in\mathbb{R}^{p}$, where $p$ is the
number of parameters. For a fixed parameter $\beta$, the dynamics
of the controlled state process $X^{c}=\left(W_{t},R_{t},Q_{t}\right)$,
valued in $\mathbb{R}^{3}$, is
\begin{eqnarray*}
X_{0}^{c} & = & x_{0}=\left(W_{0},R_{0},Q_{0}\right)\\
X_{t+1}^{c} & = & \mathcal{T}\left(X_{t}^{c},c_{t}\left(t,X_{t}^{c};\beta\right),\epsilon_{t+1}\right),t=0,...,T-1
\end{eqnarray*}
where $\left(\epsilon_{t}\right)$ is a sequence of i.i.d. random
variables, $c=c_{t}(t,X_{t}^{c};\beta)$ is the consumption policy
and $\mathcal{T}$ is a known transition function given by equations
\eqref{eq:Wt}-\eqref{eq:Rt}-\eqref{eq:Qt} and the ESG dynamics
in Table \ref{table:calibration}. 

In practice, we use Monte Carlo simulations to estimate the expectation
involved in the objective function \eqref{eq:objective}. Let $M$
be the number of Monte Carlo simulations. For each $m=1,2,\ldots,M$,
we compute a simulation path
\begin{eqnarray*}
X_{0}^{c,m} & = & x_{0}=\left(W_{0},R_{0},Q_{0}\right)\\
X_{t+1}^{c,m} & = & \mathcal{T}\left(X_{t}^{c,m},c_{t}\left(t,X_{t}^{c,m};\beta\right),\epsilon_{t+1}^{m}\right),t=0,...,T-1
\end{eqnarray*}
where $\epsilon_{t}^{m}$, $m=1,2,\ldots,M$, are i.i.d. realisations
of the random variable $\epsilon_{t}$. The counterpart of problem
\eqref{eq:objective} with sample averaging and parametric control
is given by
\begin{equation}
\hat{V}_{0}^{M}=\max_{\beta\in\mathbb{R}^{p}}\frac{1}{M}\sum_{m=1}^{M}\left[\sum_{t=0}^{T}\left\{ _{t}p_{x}u\left(c_{t}\left(t,X_{t}^{c,m};\beta\right)\right)+_{t-1|}q_{x}v\left(w_{t}^{c,m}\right)\right\} \right]\label{eq:MDUF_Monte_Carlo}
\end{equation}

\subsection{Deep neural network consumption policy}

In order to solve the empirical problem \eqref{eq:MDUF_Monte_Carlo}
in practice, we still need to choose a specific class of parametric
functions $(t,x)\in\mathbb{R}^{d+1}\shortrightarrow c\left(t,x;\beta\right)$
to model the consumption policy. We choose to model the control policy
by a deep neural network (DNN, \citealt{Goodfellow2016}). In other
words, we consider functions $c$ defined as a composition of linear
combinations and nonlinear activation functions:
\begin{align}
c\left(t,W,R,Q;\beta\right) & =\left(W+A_{t}\left(W,Q\right)\right)\times\mathcal{S}_{\mathrm{out}}\left(b^{(3)}+\sum_{k=1}^{K_{3}}w_{k}^{(3)}\varphi_{k}^{(2)}\right)\label{eq:output_layer}\\
\varphi_{k}^{(2)} & =\mathcal{S}_{2}\left(b_{k}^{(2)}+\sum_{k_{2}=1}^{K_{2}}w_{k_{2},k}^{(2)}\varphi_{k_{2}}^{(1)}\right)\,,\,\,k=1,2,\ldots,K_{3}\label{eq:2nd_hidden_layer}\\
\varphi_{k}^{(1)} & =\mathcal{S}_{1}\left(b_{k}^{(1)}+\sum_{k_{1}=1}^{K_{1}}w_{k_{1},k}^{(1)}\varphi_{k_{1}}^{(0)}\right)\,,\,\,k=1,2,\ldots,K_{2}\label{eq:1st_hidden_layer}\\
\varphi_{k}^{(0)} & =\mathcal{S}_{\mathrm{in}}\left(b_{k}^{(0)}+w_{0,k}^{(0)}t+w_{1,k}^{(0)}W+w_{2,k}^{(0)}R+w_{3,k}^{(0)}Q\right)\,,\,\,k=1,2,\ldots,K_{1}\label{eq:input_layer}
\end{align}
for any input values $(t,W,R,Q)$. Equations \eqref{eq:output_layer}-\eqref{eq:2nd_hidden_layer}-\eqref{eq:1st_hidden_layer}-\eqref{eq:input_layer}
give an example of fully connected deep neural network with four layers,
including two hidden layers. The input layer \eqref{eq:input_layer}
takes $(t,x)$ as inputs are returns $K_{1}$ ``neurons'' $\varphi_{k}^{(0)}$,
referring to nonlinear transforms of linear combinations of the inputs.
The nonlinear transformation is performed by the activation function
$\mathcal{S}_{\mathrm{in}}$. Then, the first hidden layer \eqref{eq:1st_hidden_layer}
is obtained by a linear combination of the input neurons composed
with the activation function $S_{1}$. In a similar manner, the second
hidden layer \eqref{eq:2nd_hidden_layer} is obtained by a linear
combination of the neurons \eqref{eq:1st_hidden_layer} composed with
the activation function $S_{2}$. Finally, the output layer \eqref{eq:output_layer}
is obtained by a linear combination of the neurons \eqref{eq:2nd_hidden_layer}
composed with the output activation function $\mathcal{S}_{\mathrm{out}}$.
We choose a sigmoid function $\mathcal{S}_{\mathrm{out}}(x)=1/(1+e^{-x})$
for the output activation function, scaled by the factor $\left(W+A_{t}\left(W,Q\right)\right)$
to enforce the consumption constraint \eqref{eq:consumption_constraint}.
For the other activation functions, we choose Rectified Linear Unit
(ReLU) activation functions $\mathcal{S}_{\mathrm{in}}(x)=\mathcal{S}_{1}(x)=\mathcal{S}_{2}(x)=\max(0,x)$.

The set of parameters $\beta$ contains all the weights $w^{(\ell)}$
and biases $b^{(\ell)}$, $\ell=0,\ldots3$
\[
\beta=\left\{ \left(b_{k_{1}}^{(0)},w_{k_{0},k_{1}}^{(0)}\right)_{k_{0}=0,\ldots,3}^{k_{1}=1,\ldots,K_{1}}\!\!,\left(b_{k_{2}}^{(1)},w_{k_{1},k_{2}}^{(1)}\right)_{k_{1}=1,\ldots,K_{1}}^{k_{2}=1,\ldots,K_{2}}\!,\left(b_{k_{3}}^{(2)},w_{k_{2},k_{3}}^{(2)}\right)_{k_{2}=1,\ldots,K_{2}}^{k_{3}=1,\ldots,K_{3}}\!,\left(b^{(3)},w_{k_{3}}^{(3)}\right)_{k_{3}=1,\ldots,K_{3}}\right\} 
\]
In order to train the parameters (i.e. finding the optimal parameters
$\beta^{*}$ maximising equation \eqref{eq:MDUF_Monte_Carlo}), we
perform a gradient descent using the Adam optimiser (Adaptive Moment
Estimation, \citealt{Kingma2014}), with parameter initialisation
of \citet{He2015}, using the Python machine learning library PyTorch
\citep{Paszke2019}, which takes care of gradient computations by
automatic differentiation \citep{Paszke2017}.

In practice, we consider a global consumption DNN with four layers
as described by equations \eqref{eq:output_layer}-\eqref{eq:2nd_hidden_layer}-\eqref{eq:1st_hidden_layer}-\eqref{eq:input_layer}.
There are four input dimensions (time $t$, wealth $W$, return $R$
and inflation discount factor $Q$) and one output (consumption).
We choose $K_{1}=K_{2}=K_{3}=20$ neurons for every layer, set the
optimiser learning rate to $5\times10^{-4}$ and the maximum number
of iterations to $100,000$. Finally, we produce independent training
sets and testing sets containing $M=100,000$ simulations using the
ESG described in subsection \ref{sec:ESG}. We perform our numerical
tests on an Intel{\textregistered} CPU i7-7700 @ 3.60GHz\footnote{\url{https://ark.intel.com/content/www/us/en/ark/products/97128/intel-core-i7-7700-processor-8m-cache-up-to-4-20-ghz.html}}
and NVIDIA{\textregistered} GPU GeForce{\textregistered} GTX 1070\footnote{\url{https://www.nvidia.com/en-us/geforce/10-series/}},
taking advantage of PyTorch's built-in support for CUDA{\textregistered}\footnote{\url{https://pytorch.org/docs/stable/notes/cuda.html}}.

The next section describes our numerical results.

\section{Numerical results\label{sec:numerical_results}}

In this section, we demonstrate numerical results from our trained
DNN consumption policy, which estimates the optimal consumption taking
into account age, gender, wealth, inflation, portfolio returns and
the payments from means-tested Age Pension. In the numerical test,
the individual is assumed to be aged 67 in year 2020, a single male
homeowner with total wealth of A\$$500,000$ in account-based pension
and no other testable assets or financial asset. He chooses the \textit{Balanced}
investment strategy, with $70\%$ growth assets in domestic, international
equities and property, and $30\%$ in defensive assets in deposit,
domestic and equity bonds as described in subsection \ref{subsec:Portfolio}.
This person is entitled to the Australian mean-tested Age Pension.
The maximum age is set to 108 years, so the total time horizon is
41 years.

We use our ESG to simulate $M=100,000$ scenarios for each state variable
to train our DNN optimal consumption policy. We simulate an independent
testing set of $M$ scenarios to compute the realised wealth trajectories,
Age Pension eligibility and realised lifetime utility (equation \eqref{eq:MDUF_Monte_Carlo})
from the trained DNN consumption policy. We compare the consumption
policy from our DNN approach with other alternative drawdown strategies
using the lifetime utility with risk aversion $\rho=5$ and bequest
motive $\phi=0.5$. We also conduct sensitivity analysis for retirees
who are less risk averse ($\rho=2$), and those who have no bequest
motive ($\phi=0$). We also show how this policy performs with different
initial starting wealth. Finally, we illustrate how the difference
in mortality rates between males and females affect their respective
optimal consumption policies.

\subsection{Dynamic decision making}

\begin{figure}[H]
\centering{}\includegraphics[width=12cm]{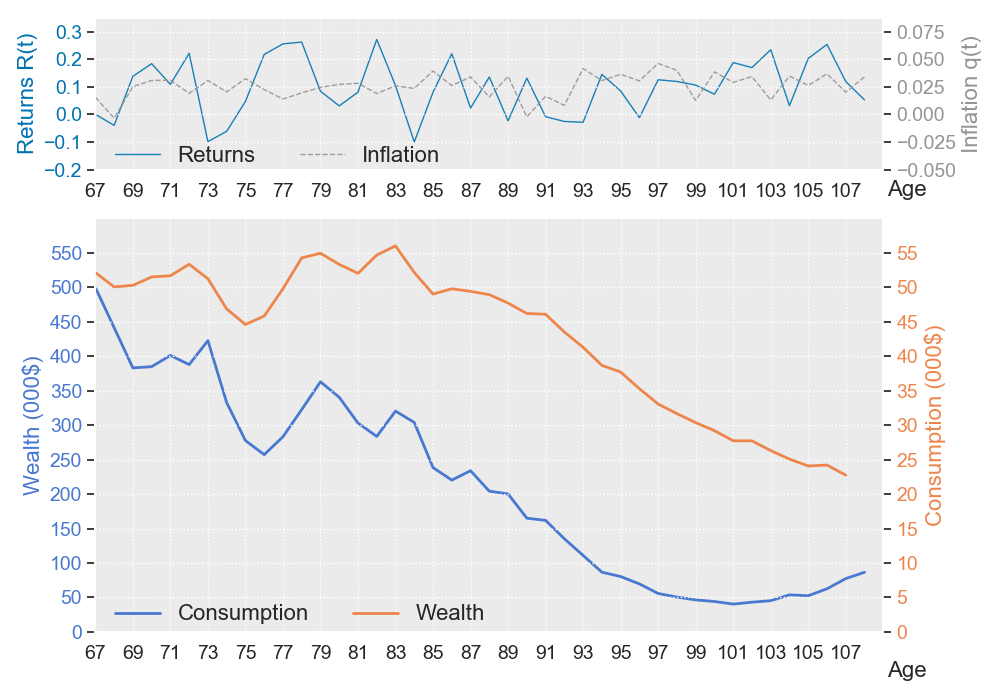}\caption{One realisation of consumption and wealth under one simulated inflation
and portfolio return with DNN optimal consumption policy.\label{fig:onePaths}}
\end{figure}

At time $t$, the trained DNN consumption policy estimates the optimal
consumption level based on the available information from simulated
inflation, portfolio returns, wealth and the Age Pension eligibility
which depends on the wealth and compound inflation before time $t$.
In Figure \ref{fig:onePaths}, we randomly pick one realisation to
demonstrate how our trained DNN policy works in the face of changes
in inflation and portfolio returns. The upper subplot displays the
simulated portfolio returns and inflation, while the lower subplot
shows the consumption suggested by the DNN policy and the resulting
wealth from the retirement age 67 to age 108.

The result shows that the optimal consumption at retirement age 67
is $\$51,917$ with partial Age Pension payment, the resulting wealth
balance at age 68 is $\$440,506$ after being adjusted for inflation
according to Eq. \ref{eq:Wt}. At age 73 and 84, the huge drops in
the market returns $R_{t}$ have impact on the wealth and consumption
at age 74 and 85. The consumption decreases over time as mortality
rate increases. At age 107, the adjusted consumption is only $\$22,900$,
that leads to a legacy wealth of $\$85,929$ at age 108. Note that
after age 67, the wealth at time $t>0$ also represents the legacy
residual if this person die between $t$ and $t+1$.

\subsection{Comparison with deterministic strategies}

With the lifetime utility measure with risk aversion $\rho=5$ and
bequest motive $\theta=0.5$, we compare our dynamic DNN consumption
strategy with six alternative deterministic drawdown strategies discussed
in \citet{Chen2020} which are: (1) the mandated age-related \textit{minimum}
drawdown rules\footnote{The minimum withdrawal rate is available at \url{https://www.ato.gov.au/rates/key-superannuation-rates-and-thresholds/?page=10}};
(2) \textit{4\%} of the initial balance in real term \citep{Bengen1994};
(3) the Rule of Thumb (\textit{RoT}), which is the minimum of the
first digit of the age of the individual as the drawdown rate, plus
2\% if the wealth is between A\$250,000 and A\$500,000 \citep{Bell2017,Deravin2019};
(4) the Association of Super Funds Australia\textquoteright s \citep{Asfa2020}\textit{
Modest} $\$28,220$ and (5) \textit{Comfortable} $\$44,183$ lifestyle
as of Mar 2020; and (6) a \textit{Luxury} consumption of $\$50,000$
per year.

\begin{figure}[H]
\centering{}\includegraphics[width=14cm]{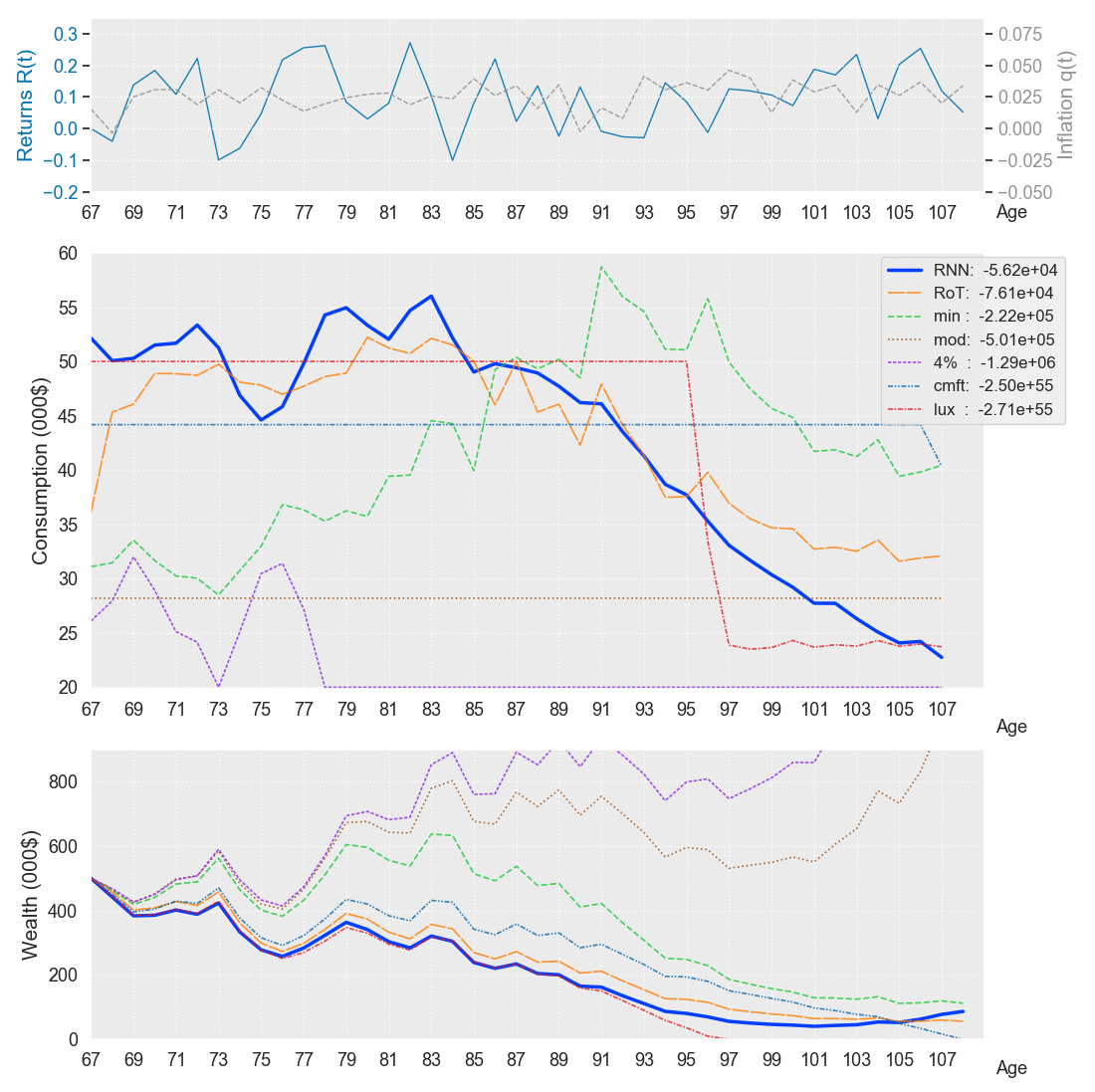}\caption{The simulated consumption and wealth paths under seven different drawdown
strategies for one realisation of inflation and portfolio returns.\label{fig:7DD}}
\end{figure}

We demonstrate such a comparison using one simulated path in Figure
\ref{fig:7DD}. The realised lifetime utilities for this path under
DNN, \textit{RoT}, \textit{minimum},\textit{ modest},\textit{ 4\%},\textit{
comfortable }and\textit{ luxury }drawdown strategies are: $U_{\mathrm{DNN}}=-5.62\times10^{4}$,
$U_{\mathrm{RoT}}=-7.61\times10^{4}$, $U_{\mathrm{min}}=-2.22\times10^{5}$,
$U_{\mathrm{mod}}=-5.01\times10^{5}$,$U_{4\mathrm{pct}}=-1.29\times10^{6}$,
$U_{\mathrm{cmft}}=-2.5\times10^{55}$, $U_{\mathrm{lux}}=-2.71\times10^{55}$
respectively. By definition, the realised lifetime utilities are negative,
and the higher (i.e. the closer to zero) the better. For this particular
scenario, the DNN strategy outperforms all the other strategies, yielding
the highest lifetime utility. The \textit{RoT} strategies, which is
a simple rule developed under similar lifetime utility framework,
takes the second place, followed by the mandated \textit{minimum}
and \textit{modest} target drawdown strategies. Using \textit{4\%}
drawdown rule with an annual consumption of $\$20,000$ in real term,
the remaining wealth keeps growing. It is not optimal as such a consumption
level is so low that the accumulated wealth makes this person no longer
eligible to receive any Age Pension payment after age 79 due to the
asset test, and this person has to live solely on the 4\% withdrawal
from the account-based pension. The \textit{comfortable} and \textit{luxury}
target consumption strategies are not sustainable under this utility
measure, as the wealth runs out at age 106 and 96 respectively.

\begin{figure}[H]
\begin{centering}
\includegraphics[width=14cm]{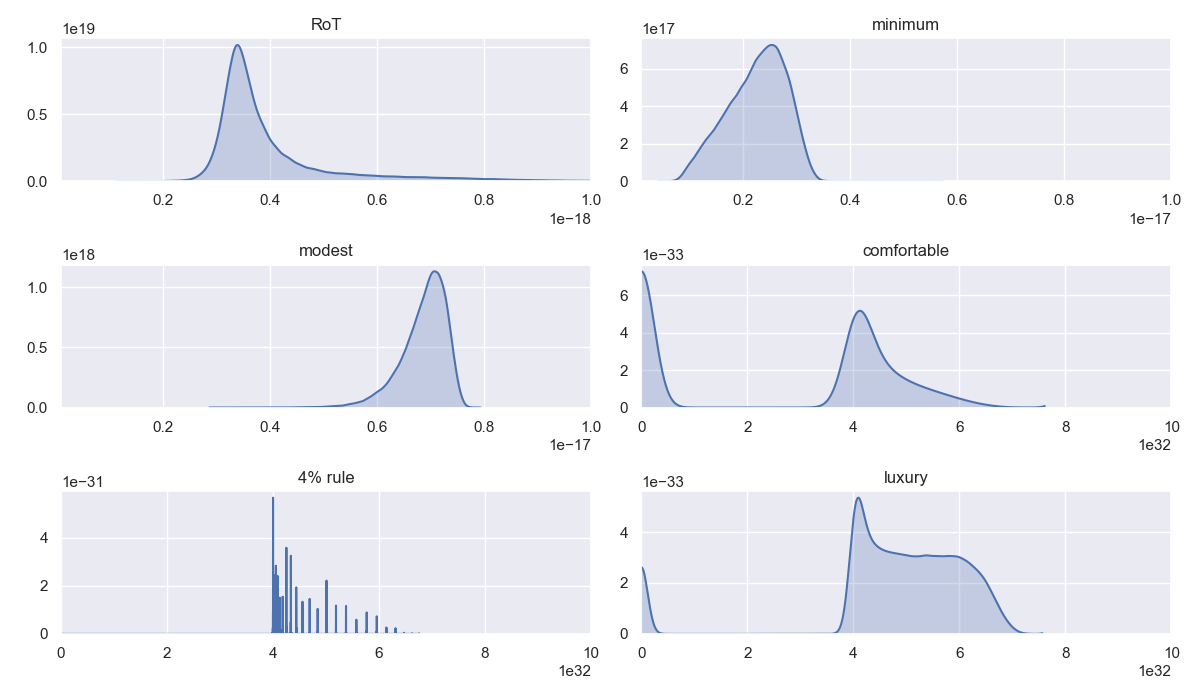}
\par\end{centering}
\centering{}\caption{The kernel density estimate of the utility differences between the
DNN drawdown and six deterministic drawdown strategies.\label{fig:u_kde}}
\end{figure}

To investigate the utility differences across the $100,000$ test
simulations, we plot the kernel density estimate of the utility differences
$U_{\mathrm{DNN}}-U_{i}$ with $i\in\left\{ \mathrm{RoT},\mathrm{min},\mathrm{mod},\mathrm{cmft},4\mathrm{pct},\mathrm{lux}\right\} $
between the DNN policy and the six deterministic drawdown strategies
in Figure \ref{fig:u_kde}. The horizontal axis of each subplot is
log-scaled. The DNN policy performs better than all these strategies.
The \textit{RoT} strategy yields the least utility difference and
performs relatively better than the rest of the fives strategies.

\begin{figure}
\centering{}\includegraphics[width=12cm]{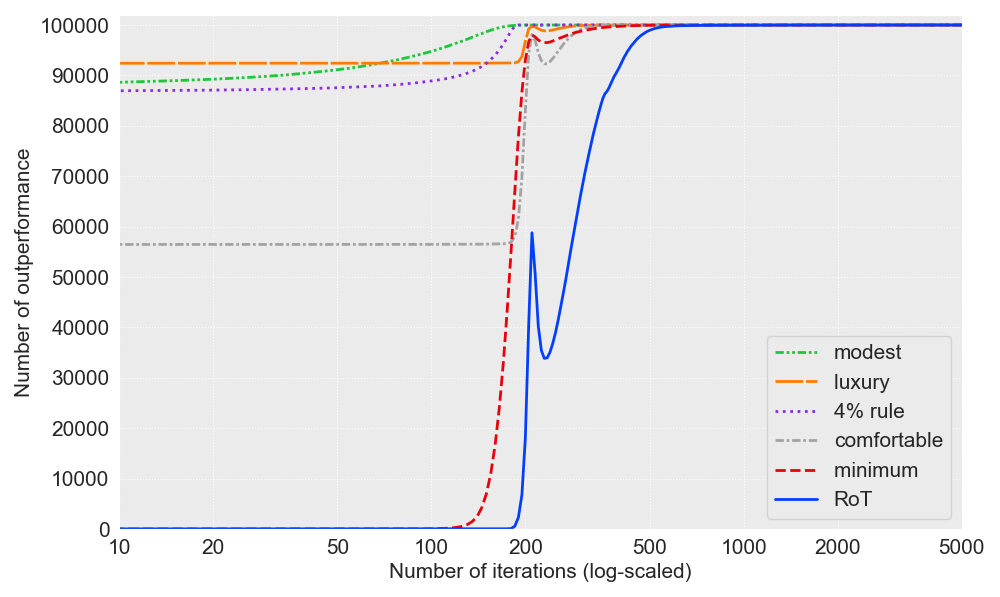}\caption{Number of DNN outperformance paths over the number of training iterations
(in log-scale).\label{fig:outperform}}
\end{figure}

In addition, the performance of the DNN policy can also be determined
with respect to the number of training iterations. In order to minimise
the loss function \eqref{eq:MDUF_Monte_Carlo}, the DNN policy is
trained on $M=100,000$ Monte Carlo scenarios and $100,000$ iterations.
To show the relationship between the performance of the DNN policy
and the number of training iterations, we plot on Figure \ref{fig:outperform}
the total number of testing scenarios for which the DNN consumption
policy outperforms alternative strategies versus the number of iterations.
Starting with random weights, our trained DNN policy begins to outperform
the \textit{modest}, \textit{luxury} and \textit{4\%} drawdown rules
in more than 80\% of the scenarios after a few iterations, and in
more than 90\% of the scenarios after 200 iterations (completed in
2 minutes) except for the \textit{RoT} strategy. Eventually, after
less than $1000$ iterations (completed in 10 minutes), the trained
DNN policy outperforms the other six strategies on all $100,000$
testing paths. 

\subsection{Utility parameters sensitivity}

\begin{figure}[H]
\begin{minipage}[t]{0.49\columnwidth}%
\hspace{-3mm}\includegraphics[width=8.5cm]{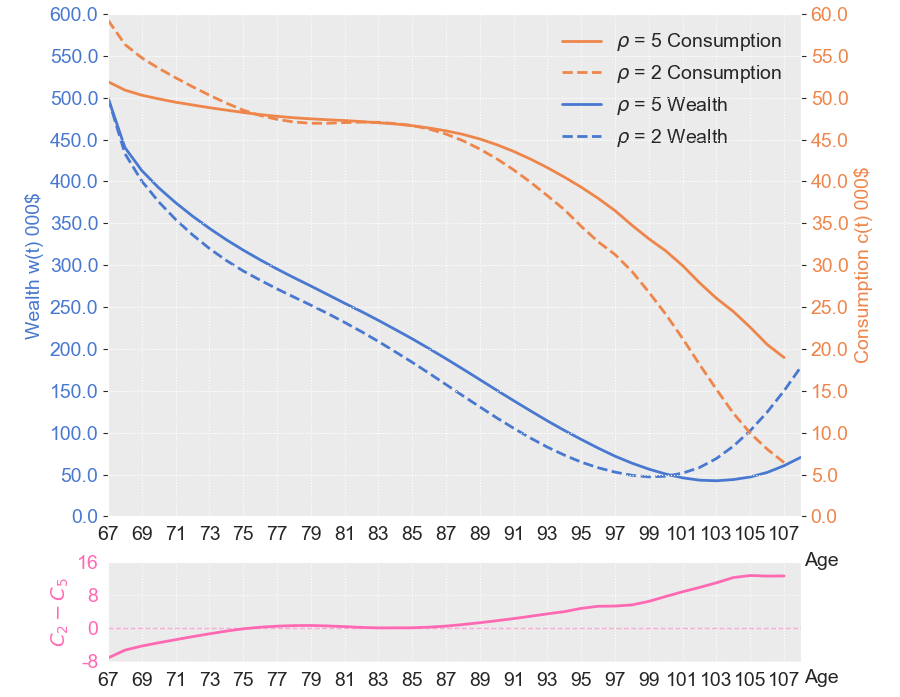}\caption{Comparison of the median consumption and median wealth with risk aversion
parameter $\rho=5$ and $\rho=2$, with bequest motive $\phi=0.5$,
and the consumption difference over a retiree's lifetime.\label{fig:risk}}
\end{minipage}$\hspace{0.5cm}$%
\begin{minipage}[t]{0.49\columnwidth}%
\includegraphics[width=8.5cm]{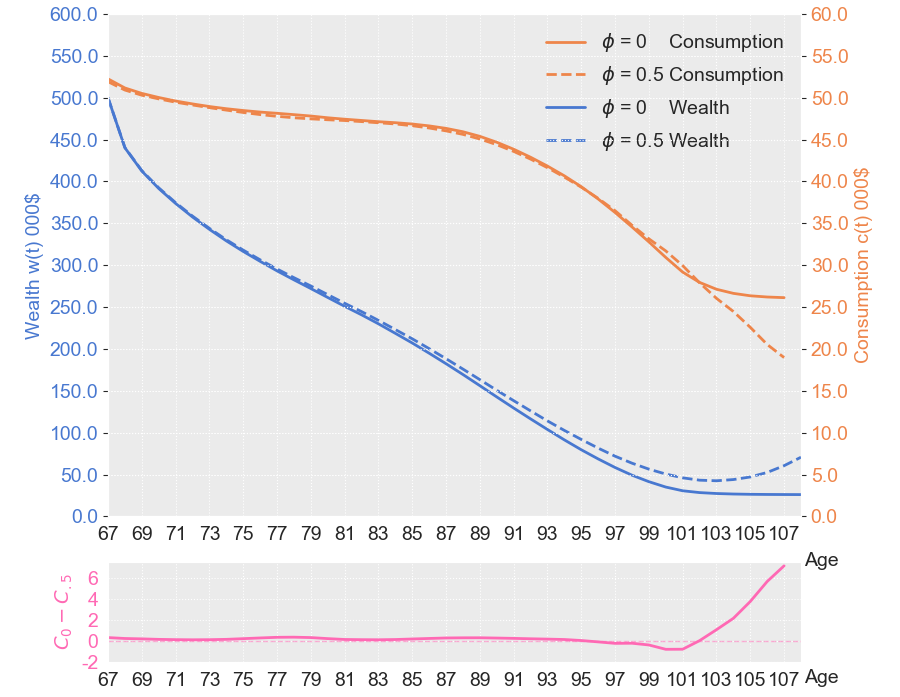}\caption{Comparison of the median consumption and median wealth with $\phi=0.5$
and without $\phi=0$ bequest motive, with risk aversion $\rho=5$,
and the consumption difference over a retiree's lifetime.\label{fig:bequest}}
\end{minipage}
\end{figure}

In this utility maximisation problem \eqref{eq:objective}, the optimal
policy also depends on the choice of utility parameters. We compare
the differences in median consumption and median wealth between two
consumption risk aversions $\rho$ and two bequest motives $\phi$
for a male retiree. We use $\rho=5$ and $\phi=0.5$ as our base case,
and conduct two analyses with lower risk aversion parameter $\rho=2$
reported in Figure \ref{fig:risk}, and a lower bequest motive $\phi=0$
reported in Figure \ref{fig:bequest}. From the upper subplot of Figure
\ref{fig:risk}, we can see that under lower risk aversion $\rho=2$,
the median of the first year optimal annual consumption (dashed orange
curve) can start from a higher level of $\$59,270$, compared to $\$51,910$
for higher risk aversion $\rho=5$ (solid orange curve). The lower
subplot of Figure \ref{fig:risk}, shows that the consumption gap
decreases first and changes of sign after age 87. To sum up, retirees
with higher consumption risk aversion tend to save more for later
consumption to avoid potential shortfalls, and their consumption curve
is flatter.

The upper subplot of Figure \ref{fig:bequest} shows the median consumption
and median wealth with ($\phi=0.5$) and without ($\phi=0$) bequest
motive. Without any bequest motive, the medians of initial consumption
$c_{0}$ is $\$52,247$, which is $\$336$ more than the optimal consumption
$\$51,910$ with bequest motive. The lower subplot of Figure \ref{fig:bequest}
shows that there is no significant difference in consumption until
age $102$ when the person can start to spend more if no desire to
leave any legacy. Such results may be counter-intuitive, as one might
think that a lot more can be spent in early retirement. However, our
results indicate that even without any bequest motive, overspending
at early retirement is not optimal. The reason is that capital returns
play a very important role in the wealth aggregation in retirement
as there is a tradeoff between consumption level and sustainability
of the wealth. Therefore, the optimal consumption is less sensitive
to the bequest motive parameter than to the risk aversion parameter.

\subsection{Consumption with different initial balances}

\begin{figure}[H]
\begin{minipage}[t]{0.5\columnwidth}%
\includegraphics[width=8.5cm]{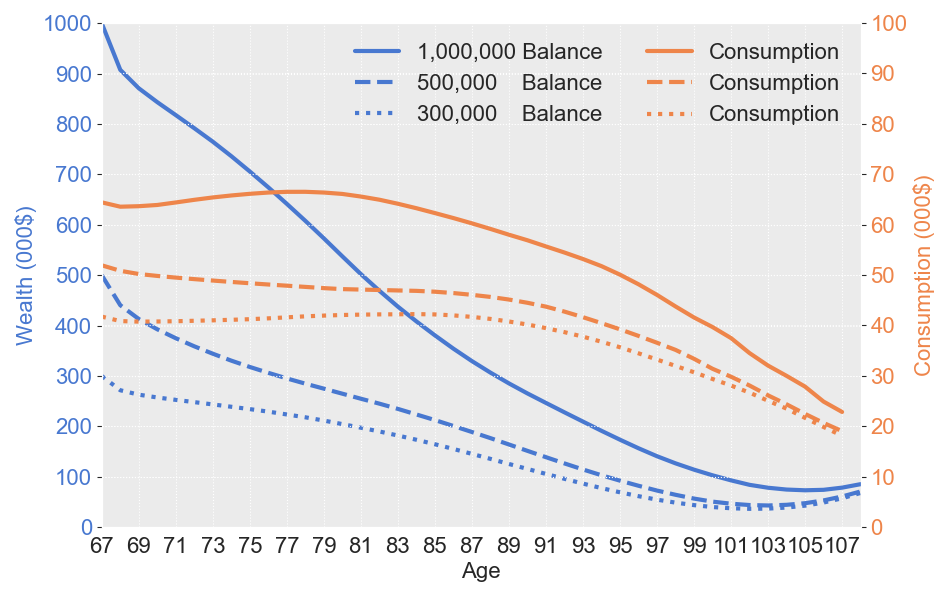}\caption{Comparison of the wealth and consumption over a retiree's lifetime
with different initial wealth $w0$.\label{fig:balance}}
\end{minipage}$\hspace{0.5cm}$%
\begin{minipage}[t]{0.46\columnwidth}%
\includegraphics[width=8cm]{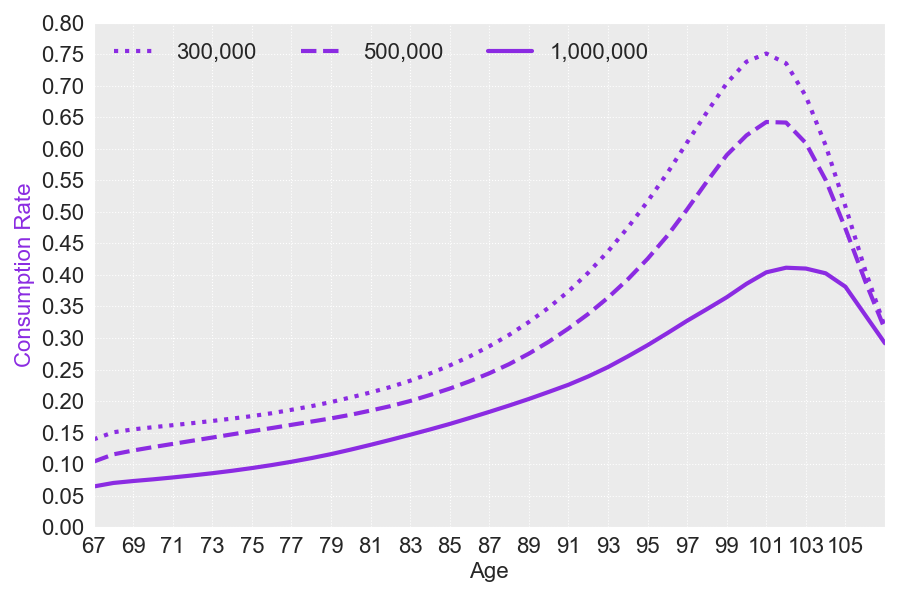}\caption{Comparison of the consumption rate over a retiree's lifetime with
different initial wealth $w0$.\label{fig:conRate}}
\end{minipage}
\end{figure}

We now investigate the effect of the initial wealth balance. We demonstrate
the medians of the simulated wealth and consumption with starting
balances of $\$300,000$, $\$500,000$, $\$1,000,000$ in Figure \ref{fig:balance}
and the consumption rates (consumption divided by wealth) in Figure
\ref{fig:conRate}. Because of the eligibility to receive full or
partial Age Pension from the government for the retirees who own asset
below the asset test limit, the retirement consumption is not proportional
to the initial wealth. Retirees with lower wealth can draw down at
a higher rate as there exist a pension \textit{safety net}.

\subsection{Gender difference in consumption}

Finally, we investigate the effect of gender on optimal consumption
decisions during retirement. The Life Table 2015-2017 from ABS shows
that the mortality of males is higher than that of females. In 2020,
the life expectancy with 25 year improvement factor for age 65 is
85.5 for male and 87.9 for female \citep{AGA2017}. To obtain realistic
measure of longevity, we consider the possible future improvements
in mortality that may occur in the future. We implement the adjustment
for mortality with the 25 year improvement factor\footnote{Details about the mortality improvement factors can be found on the
Australian Government Actuary website \url{http://www.aga.gov.au/publications/life\_table\_2015-17/default.asp}} to project the mortality that could be expected to occur over a retiree's
lifetime.

\begin{figure}[h]
\centering{}\includegraphics[width=8.5cm]{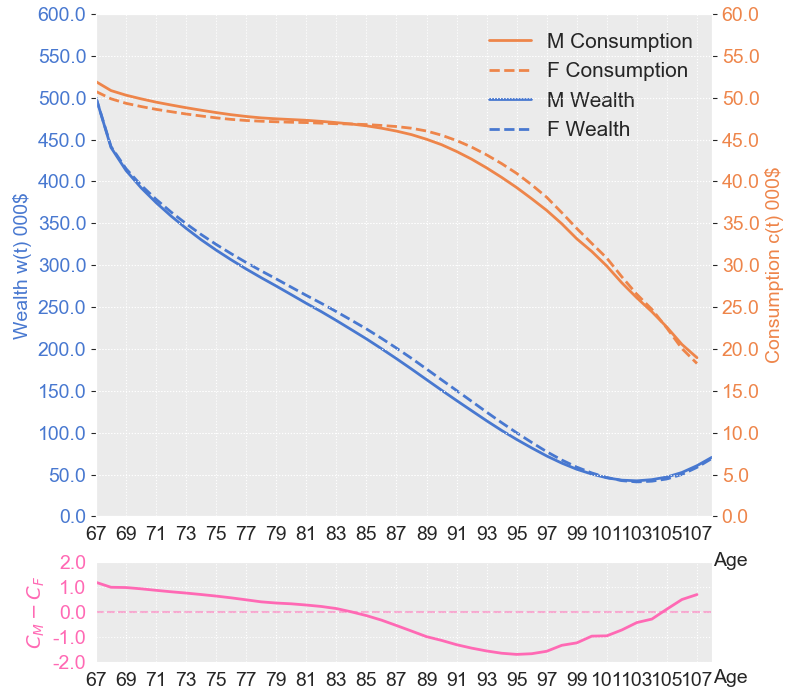}$\:$\includegraphics[width=8.5cm]{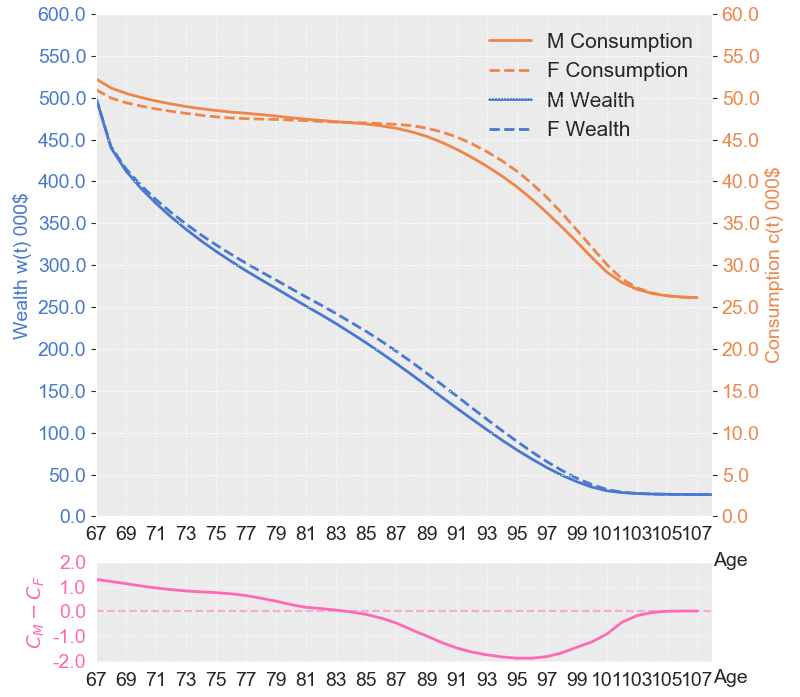}\caption{Comparison of the median consumption and median wealth between a male
and a female with ($\phi=0.5$, left) and without ($\phi=0$, right)
bequest motive under consumption risk aversion $\rho=5$.\label{fig:gender}}
\end{figure}

In Figure \ref{fig:gender}, we plot the median of the simulated optimal
consumption controlled by the DNN policy and the resulting wealth
for a male and a female aged 67 with initial balance $w_{0}=500,000$.
From Figure \ref{fig:gender}, we can observe a small effect of gender
on consumption decisions: we can see that for a 67-year-old man and
woman retiring with the same initial wealth, it is optimal for the
man to spend more than the women until age 84 either with or without
any bequest motive. Males have higher mortality rate and thus less
longevity risk than females, and therefore it is optimal for them
to consume slightly more in early retirement age. Conversely, women
have greater longevity risk and therefore should consume slightly
more conservatively than men in early retirement. Such lifetime uncertainty
increases consumption impatience which is consistent with the findings
in \citet{Huang2012} and Irving Fisher's impatience theory \citep{Fisher1930}
in behavioral economics.

\section{Conclusion\label{sec:conclusion}}

In this paper, we propose a deep neural network (DNN) optimal consumption
policy analysis for the decumulation phase in a defined contribution
(DC) pension system under a lifetime utility framework. The trained
DNN can estimate the optimal consumption level for a retiree under
any financial conditions that could happen during retirement. This
DNN approach is independent of the simulation tool used to represent
the possible future economic conditions. In practice, we propose and
use a specific seven-factor Economic Scenario Generator (ESG), sufficient
for decumulation analysis, and calibrate it to Australian market data.

We demonstrate that a DNN-based decumulation policy can successfully
be trained from a realistic multivariate economic scenario generator
in a matter of minutes. Our numerical tests demonstrate that the DNN-based
policy outperforms six common deterministic decumulation rules by
yielding higher lifetime utility for all testing scenarios after less
than $1000$ training iterations (completed in 10 minutes). We also
report the densities of lifetime utility improvement provided by the
DNN consumption policy compared to these deterministic policies. A
sensitivity analysis with respect to the risk aversion parameter $\rho$
reveals that a higher consumption risk aversion generates a smoother
and flatter consumption pattern, with less variation over time. This
contrasts with the bequest motive parameter $\phi$, which is shown
to have a negligible effect on optimal consumption until very late
in retirement. Due to the safety net provided by the means-tested
Age Pension payment, we observe that the optimal consumption is not
proportional to the wealth of the retiree. Finally, gender has a small
effect on optimal consumption due to the difference in mortality between
males and females. Our results suggest that females should consume
slightly less in early retirement (before age 84) to mitigate their
greater longevity risk.

The DNN-policy approach developed in this paper for Australia can
be applied to similar DC systems in other countries (e.g. UK DC system,
USA 401(k)) with the necessary country adjustments and taking into
account the government funded pension payment policy. Moreover, it
can be extended to incorporate other financial decisions such as asset
allocation or partial annuitisation, which we leave for future research.

\bibliographystyle{apalike}
\bibliography{biblio}

\appendix

\section{Calibration of ESG parameters\label{sec:Calibration}}

This Appendix details the historical data used to calibrate the ESG
and provides the calibrated parameter values in order to make our
results replicable. The ESG parameters are calibrated by ordinary
least squares method using historical data from 1992, the year when
the compulsory contribution system started, to 2020 as shown in Table
\ref{tab:historical}. The short-term interest rates $s(t)$ are the
90-day bank accepted bills from the RBA website. Most data are indices,
except for the short-term interest rates $s(t)$, therefore we first
compute the returns using the log-ratio of two consecutive indices
in time. The inflation rate $q(t)=\log\left(\text{CPI (t)}/\text{CPI (t-1)}\right)$
is the log-ratio of the all groups Consumer Prices Index $\text{CPI}(t)$
from the RBA. The domestic $e(t)$ and international equity total
returns $n(t)$ are computed from the S\&P/ASX200 Accumulation Index
$E(t)$ and MSCI World ex-Australia Net Index (in Australian Dollar)
$N(t)$ respectively. While the domestic $b(t)$ and international
bond returns $o(t)$ are computed from the AusBond Index $B(t)$ and
Citigroup World Government Bond Index $O(t)$. These four indices
are downloaded from Bloomberg. The house price growth $h(t)$ comes
from the House Price Index $\text{HPI}(t)$ which is the Established
House Price Index (weighted average of 8 capital cities) from the
ABS website\footnote{The ABS has compiled a House Price Index (HPI) since 1986. A significant
review of the HPI occurred in 2004 and a new series of HPI was introduced
in 2005. We scaled the HPI before 2005 to keep the house price returns
consistent with the data before 2005. More details about the HPI is
available at \url{https://www.abs.gov.au/ausstats/abs@.nsf/mf/6416.0}}.

\begin{table}
\centering{}%
\begin{tabular}{l|lllllll}
\hline 
Year & $\text{CPI}(t)$ & $s(t)$ $\%$ & $E(t)$ & $N(t)$ & $B(t)$ & $O(t)$ & $\text{HPI}(t)$\tabularnewline
\hline 
1992 & \enskip{}59.7 & 6.42 & \enskip{}5808.3 & \enskip{}902.2 & \enskip{}1692.5 & \enskip{}299.6 & \enskip{}28.73\tabularnewline
1993 & \enskip{}60.8 & 5.25 & \enskip{}6291.6 & 1602.4 & \enskip{}1928.3 & \enskip{}373.1 & \enskip{}29.50\tabularnewline
1994 & \enskip{}61.9 & 5.47 & \enskip{}7252.1 & 1591.3 & \enskip{}1906.5 & \enskip{}357.9 & \enskip{}30.49\tabularnewline
1995 & \enskip{}64.7 & 7.57 & \enskip{}7756.2 & 1817.0 & \enskip{}2133.0 & \enskip{}438.2 & \enskip{}30.90\tabularnewline
1996 & \enskip{}66.7 & 7.59 & \enskip{}8866.4 & 1937.9 & \enskip{}2334.6 & \enskip{}396.1 & \enskip{}31.23\tabularnewline
1997 & \enskip{}66.9 & 5.28 & 11313.3 & 2491.4 & \enskip{}2725.8 & \enskip{}432.3 & \enskip{}32.11\tabularnewline
1998 & \enskip{}67.4 & 5.32 & 11542.2 & 3541.8 & \enskip{}3022.4 & \enskip{}545.5 & \enskip{}34.89\tabularnewline
1999 & \enskip{}68.1 & 4.93 & 13251.6 & 3830.8 & \enskip{}3121.7 & \enskip{}531.5 & \enskip{}36.92\tabularnewline
2000 & \enskip{}70.2 & 6.23 & 15628.0 & 4742.7 & \enskip{}3314.3 & \enskip{}604.5 & \enskip{}40.50\tabularnewline
2001 & \enskip{}74.5 & 4.97 & 17044.8 & 4457.9 & \enskip{}3559.7 & \enskip{}592.3 & \enskip{}43.82\tabularnewline
2002 & \enskip{}76.6 & 5.07 & 16245.3 & 3410.4 & \enskip{}3781.3 & \enskip{}714.1 & \enskip{}52.1\tabularnewline
2003 & \enskip{}78.6 & 4.67 & 15966.7 & 2778.2 & \enskip{}4151.3 & \enskip{}696.6 & \enskip{}61.2\tabularnewline
2004 & \enskip{}80.6 & 5.49 & 19416.7 & 3316.6 & \enskip{}4247.9 & \enskip{}712.2 & \enskip{}70.9\tabularnewline
2005 & \enskip{}82.6 & 5.66 & 24533.9 & 3318.6 & \enskip{}4578.8 & \enskip{}641.5 & \enskip{}71.0\tabularnewline
2006 & \enskip{}85.9 & 5.96 & 30405.1 & 3978.3 & \enskip{}4735.1 & \enskip{}712.2 & \enskip{}73.8\tabularnewline
2007 & \enskip{}87.7 & 6.42 & 39119.1 & 4287.3 & \enskip{}4923.0 & \enskip{}641.5 & \enskip{}80.9\tabularnewline
2008 & \enskip{}91.6 & 7.81 & 33875.3 & 3375.9 & \enskip{}5141.6 & \enskip{}663.5 & \enskip{}91.8\tabularnewline
2009 & \enskip{}92.9 & 3.25 & 27053.6 & 2827.8 & \enskip{}5698.0 & \enskip{}819.1 & \enskip{}86.8\tabularnewline
2010 & \enskip{}95.8 & 4.89 & 30610.0 & 2975.3 & \enskip{}6145.6 & \enskip{}807.7 & 103.1\tabularnewline
2011 & \enskip{}99.2 & 4.99 & 34200.7 & 3054.4 & \enskip{}6486.5 & \enskip{}704.4 & 103.2\tabularnewline
2012 & 100.4 & 3.49 & 31904.5 & 3039.0 & \enskip{}7290.8 & \enskip{}755.4 & \enskip{}99.7\tabularnewline
2013 & 102.8 & 2.80 & 39163.3 & 4045.1 & \enskip{}7492.6 & \enskip{}807.9 & 103.1\tabularnewline
2014 & 105.9 & 2.70 & 45991.2 & 4870.5 & \enskip{}7950.1 & \enskip{}837.1 & 114.4\tabularnewline
2015 & 107.5 & 2.15 & 48602.3 & 4686.0 & \enskip{}8397.5 & \enskip{}935.2 & 123.0\tabularnewline
2016 & 108.6 & 1.99 & 48872.4 & 4557.9 & \enskip{}8986.8 & 1074.1 & 132.1\tabularnewline
2017 & 110.7 & 1.72 & 55758.6 & 5386.9 & \enskip{}9009.23 & \enskip{}999.5 & 147.3\tabularnewline
2018 & 113.5 & 2.07 & 63015.4 & 5987.6 & \enskip{}9287.2 & 1057.4 & 150.6\tabularnewline
2019 & 114.8 & 1.29 & 70291.8 & 6636.4 & 10176.2 & 1174.3 & 138.1\tabularnewline
2020 & 116.6 & 0.102 & 64116.9 & 6583.0 & 10601.6 & 1262.5 & 150.3\tabularnewline
\hline 
\end{tabular} \caption{Australian historical data from June 1992 to June 2020.\label{tab:historical}}
\end{table}

\begin{table}
\centering{}%
\begin{tabular}{lcl}
\hline 
Variable dynamics & Params & Values\tabularnewline
\hline 
1, Price inflation $q(t)$ & $\mu_{q}$ & \multicolumn{1}{l}{\enskip{}$0.024$}\tabularnewline
\multirow{2}{*}{$q(t)=(1-\phi_{q})\mu_{q}+\phi_{q}q(t-1)+\epsilon_{q}(t)$} & $\phi_{q}$ & \enskip{}$0.1346$\tabularnewline
 & $\sigma_{q}$ & \enskip{}$0.012$\tabularnewline
\hline 
\multicolumn{1}{l}{2, Short-term interest rate $s(t)$} & $\text{\ensuremath{\mu}}_{S}$ & \enskip{}$0.141$\tabularnewline
\multicolumn{1}{l}{$S(t)=\phi_{S}S(t-1)+\left(1-\phi_{S}\right)\left(\mu_{S}-\mu_{q}\right)+\epsilon_{s}(t)$} & $\phi_{S}$ & \enskip{}$0.813$\tabularnewline
$s(t)=S(t)+q(t)$ & $\sigma_{S}$ & \enskip{}$0.015$\tabularnewline
\hline 
\multicolumn{1}{l}{3, Domestic equity returns $e(t)$} & $\mu_{e}$ & \enskip{}$0.085$\tabularnewline
$e(t)=\phi_{e}e(t-1)+(1-\phi_{e})\mu_{e}+\epsilon_{e}$ & $\phi_{e}$ & \enskip{}$0.164$\tabularnewline
 & $\sigma_{e}$ & \enskip{}$0.119$\tabularnewline
\hline 
\multicolumn{1}{l}{4, International equity returns $n(t)$} & $\psi_{n,0}$ & $-0.018$\tabularnewline
$n(t)=\psi_{n,0}+\psi_{n,1}n(t-1)+\psi_{n,2}e(t)+\epsilon_{n}(t)$ & $\psi_{n,1}$ & \enskip{}$0.104$\tabularnewline
 & $\psi_{n,2}$ & \enskip{}$0.911$\tabularnewline
 & $\sigma_{n}$ & \enskip{}$0.090$\tabularnewline
\hline 
\multirow{2}{*}{5, Domestic bond $b(t)$} & $\psi_{b,0}$ & \enskip{}$0.073$\tabularnewline
 & $\psi_{b,1}$ & $-0.103$\tabularnewline
\multirow{2}{*}{$b(t)=\psi_{b,0}+\psi_{b,1}b(t-1)+\psi_{b,2}n(t)+\epsilon_{b}(t)$} & $\psi_{b,2}$ & $-0.050$\tabularnewline
 & $\sigma_{b}$ & \enskip{}$0.036$\tabularnewline
\hline 
\multirow{2}{*}{6, International bond $o(t)$} & $\psi_{o,0}$ & $-0.026$\tabularnewline
 & $\psi_{o,1}$ & \enskip{}$1.340$\tabularnewline
\multirow{2}{*}{$o(t)=\psi_{o,0}+\psi_{o,1}e(t)+\psi_{o,2}n(t)+\epsilon_{o}(t)$} & $\psi_{o,2}$ & $-0.200$\tabularnewline
 & $\sigma_{o}$ & \enskip{}$0.081$\tabularnewline
\hline 
\multicolumn{1}{l}{7, House price $h(t)$} & $\psi_{h,0}$ & \enskip{}$0.066$\tabularnewline
\multirow{3}{*}{$h(t)=\psi_{h,0}+\psi_{h,1}q(t)+\psi_{h,2}b(t)+\epsilon_{h}(t)$} & $\psi_{h,1}$ & $-0.489$\tabularnewline
 & $\psi_{h,2}$ & \enskip{}$1.037$\tabularnewline
 & $\sigma_{h}$ & \enskip{}$0.061$\tabularnewline
\hline 
\end{tabular}\caption{Dynamics of the variables and calibrated parameters of the ESG\label{table:calibration}}
\end{table}

\begin{figure}

\centering{}\includegraphics[width=12cm]{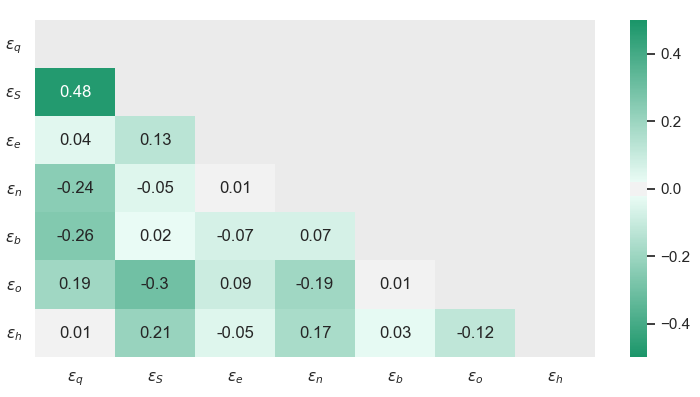}\caption{Cross-correlation matrix of the residuals.\label{fig:corr}}
\end{figure}

We assume that the residuals are conditionally independent. Figure
\ref{fig:corr} shows the lower triangular part of their correlation
matrix which exhibit low correlations as expected.

\begin{figure}
\centering{}\includegraphics[width=14cm]{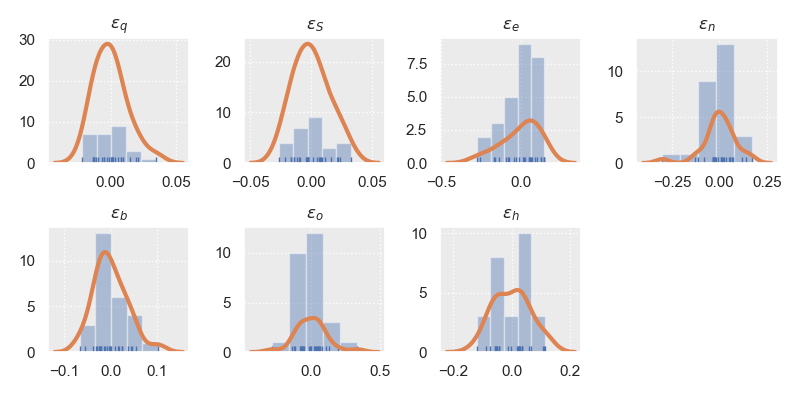}\caption{Kernel density estimate of empirical errors.\label{fig:kde}}
\end{figure}

We plot the histogram and the kernel density estimate of the empirical
errors. The empirical distributions of the residuals look not too
far from being centered, unimodal, symmetric, and can be approximated
by normal variables.
\end{document}